\documentclass[symmetry, article, submit, pdftex, oneauthor]{mdpi} 

\firstpage{1}
\pubvolume{1}
\issuenum{1}
\articlenumber{0}
\pubyear{2025}
\copyrightyear{2025}
\datereceived{ }
\daterevised{ } % Comment out if no revised date
\dateaccepted{ }
\datepublished{ }
\hreflink{https://doi.org/} % If needed use \linebreak
\preto{\abstractkeywords}{\nolinenumbers}

\usepackage{graphicx,amsfonts,amssymb,amsmath,array,xcolor}
\usepackage[utf8]{inputenc}

\newlength{\ldag}
\Title{The power spectrum of compressible turbulence: A story in symmetries}

\TitleCitation{The power spectrum of compressible turbulence: A story in symmetries}

\Author{Olivier Coquand $^{1}$\orcidA{}}

\AuthorNames{Olivier Coquand}

\isAPAStyle{%
       \AuthorCitation{Coquand, Olivier}
         }{%
        \isChicagoStyle{%
        \AuthorCitation{Coquand, Olivier}
        }{
        \AuthorCitation{Coquand, Olivier}
        }
}

\address[1]{%
$^{1}$ \quad Laboratoire de Modélisation Pluridisciplinaire et Simulations, Université de Perpignan Via Domitia,
52 avenue Paul Alduy, F-88860 Perpignan, France; olivier.coquand@univ-perp.fr}

\corres{Correspondence: olivier.coquand@univ-perp.fr}

\abstract{This article presents a theoretical study of the scaling properties of the kinetic energy spectrum in compressible turbulence.
From the fundamental symmetries and linear transformations of the microscopic action, we derive exact relations between the correlation functions
and their generators.
These relations put strong constraints on the possible scaling relations in the system as a function of scale.
One of the main results of this study is that the action can be split between an incompressible part, which is the same as in the usual
stochastic Navier-Stokes theory whatever the value of the Mach number, and a longitudinal part, whose behavior is to be compared to the three
dimensional Burgers equation, which presents a much richer phase diagram as its usually discussed one dimensionalcounterpart.}


\begin{document}

\section{Introduction}

	Turbulence in Newtonian fluids is certainly one of the oldest problems in out-of-equilibrium physics that is still today under
	active investigation.
	One of the biggest breakthrough in its study has certainly been the work of Kolmogorov in 1941 \cite{Kolmogorov41a, Kolmogorov41b, Kolmogorov41c},
	who has shown that, although the motion of the turbulent fluid seems erratic, there exist some universally quantifiable regularity in the behaviour
	of some of its statistical properties, like the velocity pair correlation function, at least under some assumptions of isotropy, average homogeneity,
	and, more crucially, incompressibility of the fluid.

	The case of compressible fluids though, is also of great importance in various fields of physics, mostly for applications in aeronautics, but also in
	astrophysics as it is the basis of models of the interstellar medium, the generation of galaxies and stars and their evolution through time.
	The scaling of statistical characteristics of compressible fluids however remains still largely elusive today, as witnessed by evaluations of the
	situation in relatively recent papers: "the theoretical understanding of compressible turbulence is still poor" \cite{Konstandin11},
	"To our knowledge, no universal law has been derived for compressible turbulence" \cite{Galtier11}, or "The particle distribution function and spectrum
	is still a matter of debate" \cite{Federrath13}.

	Most of the studies of the subject have been conducted numerically. They do not yield a clear and consistent picture
	of how compressible fluids behave in turbulent flows.
	It is to be noted that along the years, numerical methods have evolved, which surely explains part of the dispersion of results.
	There are also two types of limitations that are intrinsic to numerical simulations that can explain at least partly the discrepancies with
	the theoretical results --- exposed hereafter --- first the fact that simulations are conducted on the Euler equation and thus the viscosity is
	generated numerically rather than given as a physical input in the equations of motion, and second the so-called bottleneck effect that reduces
	the effective size of the inertial range in which universal behaviors are to be expected \cite{Federrath13}.

	Without being exhaustive, the recent numerical results regarding the scaling laws of compressible turbulence can be summarized as follows:
	In \cite{Kritsuk07}, it is exposed that the statistical characteristics of the flows stay similar to those of incompressible fluids for Mach
	numbers Ma$\lesssim1$. Then, the study conducted for Ma$=6$ regarding the scaling of the power law exponent of the kinetic
	energy spectrum $\beta$ yields $\beta = -1.95$, in contrast to the $-5/3$ value predicted by Kolmogorov in the incompressible case (hereafter
	denoted K41 scaling for short).
	The density of the fluid is predicted to scale as $\rho(k)\sim k^{0.45}$.

	In \cite{Schmidt08}, the influence of the type of forcing on supersonic flows is examined; In particular, the extreme limits of purely solenoidal or purely
	compressible forcings are investigated. It is then shown that the dependence on the type of forcing is strongly reduced if one studies the evolution of the
	mass-weighted velocity $\rho^{1/3}\mathbf{v}$ instead of the velocity $\mathbf{v}$.
	In a following paper, \cite{Konstandin11}, it is shown that this effect is exacerbated if the Eulerian framework is used in contrast to the Lagrangian one,
	at least for Mach numbers in the range Ma$\sim 4.4-4.9$.
	A subsequent study \cite{Federrath13}, conducted at higher Mach numbers Ma$=17$ computes more precisely the values of the scaling exponents in the
	case of the mass-weighted velocity distribution function.
	In particular, they report two exponents $\beta=-1.74$ for the case of solenoidal driving, and $\beta=-2.10$ in the case of compressible driving,
	showing a behavior closer to that of the 1 dimensional (1D) Burgers fluid where $\beta=-2$ (and very close to the theoretical prediction of \cite{Galtier11}
	presented below).
	In agreement with the above, the incompressible K41 scaling is recovered for Ma$\lesssim1$.

	Then, in the group of J. Wang and their collaborators, two studies at moderate Mach numbers have concluded first that the overall exponent $\beta=-2$
	\cite{Wang13}, in agreement with the 1D Burgers prediction, and, in a refined study \cite{Wang17} that in the non linear subsonic regime
	$0.5<$Ma$<1$, the velocity increment can be split into one positive component with 1D Burgers scaling, and a negative component with K41 scaling.
	On the other hand, \cite{Sun17} concluded the same year that for Ma$\gtrsim 0.3$, corrections to the 1D Burgers scaling are to be expected.
	The study gives explicit expressions for these corrections.

	Finally, \cite{Donzis20} concluded after an analysis of a large database of direct numerical simulation (DNS) of the compressible Navier-Stokes equations
	that the Reynolds number Re and Ma were not sufficient to characterize correctly the scaling properties of the fluid, since the data comprising different
	geometries and types of forcing do not collapse well onto one master curve.
	They propose instead to use $\delta$, the ratio of the root mean square parts of the incompressible and compressible parts of the velocity vector together
	with Ma.
	They show a good collapse of all the data of the ratio of the incompressible to the compressible mean kinetic energy dissipation rate as a function of $\delta$
	on a single straight line, which is an indication that Prandtl's Misschungsweghypothese \cite{Prandtl26c} holds also for compressible fluids.
	As Ma increases, $\delta$ converges towards a finite value $\delta_\infty$, a result in qualitative agreement with previous theoretical results
	exposed in \cite{Kraichnan54} and \cite{Staroselsky90}.
	Quantitative agreement between all these approaches is not reached however, probably because of a remaining lack of precision.

	On the theoretical side, one of the first studies is probably that of Kraichnan in the early 1950's \cite{Kraichnan54}, where it was already shown, by constructing
	an analogy with Hamiltonian dynamical systems, that there is an equipartition of the kinetic energy in the two incompressible velocity modes on the one
	hand, and in the two acoustic (the longitudinal velocity and compressibility) modes on the other hand.
	A little bit less than forty years later, the first renormalisation group (RG) analysis were conducted on the compressible turbulence problem \cite{Staroselsky90},
	showing that in the limit of zero viscosity, the renormalised Mach number converges towards a finite value, so that dissipation by shock waves as well
	as by momentum transfer between vortices coexist even in this limit.
	In the weakly compressible limit, it is found that the kinetic energy spectrum behaves as $E(k) = E_{inc}(k) + O(\text{Ma}^8)$ --- $E_{inc}(k)$ being
	the incompressible K41 contribution --- thus substantiating the claim that the K41 scaling holds beyond the very small Ma number values, a result consistent
	with most of the numerical studies presented above.
	In the case of a finite Mach number, it is found that the ratio of $E_{inc}(k)$ to the compressible component $E_{comp}(k)$ is a constant number, which means
	that in the supersonic case, both contribution must share the same scaling with $k$.

	In a subsequent perturbative RG computation, Antonov and his group concluded that the first correction to the behavior of the kinetic energy spectrum
	with the Mach number should rather be ${E(k)\sim \varepsilon^{2/3}k^{-5/3}\big[1+A\,\text{Ma}^2(kL)^{-2/3}\big]}$, where $\varepsilon$ is the
	mean energy dissipation rate, $A$ is a numerical constant, and $L$ is the macroscopic integral scale.
	This correction is much greater than that proposed by the group of Staroselsky.

	Finally, in \cite{Galtier11}, a re-examination of the concept of energy cascade in the context of compressible turbulent flows is proposed.
	It is found that the energy balance equation involves new terms, among which a crucial $\mathbf{r}$-dependent term that gives rise to a new scaling regime
	in the supersonic regime with $\beta=-19/9$, in close agreement with the numerical results of \cite{Federrath13}.

	To our knowledge, there is no universally accepted scenario that could solves the above exposed controversies.
	If some discrepancies can be explained by the improvement of the computing techniques --- both numerical and analytical --- across the examined period,
	the true scaling laws controlling the statistics of compressible turbulent flows, if they exist, remain hidden by a veil of mystery.

	More recently, L. Canet and her collaborators developed a RG technique that proved remarkably successful in the study of incompressible turbulence
	\cite{Canet15,Canet16,Canet17,Tarpin19,Gorbunova20,Gorbunova21a,Canet22}.
	In particular, it was shown in \cite{Canet15} that local and global symmetries of the Navier-Stokes action generated in the framework of
	Martin-Siggia-Rose-Janssen-de Dominicis (MSRJD) \cite{Martin73,Janssen76,Dominicis76} impose \textit{exact} relations between the correlation
	functions of the turbulence problems and their generating functionals.
	These exact relations put strong constraints on the possible forms of the evolution of these correlation functions with scale, and in turn, their scaling properties.

	The aim of the present work is to adapt this formalism to the case of compressible flows, to investigate what predictions can be deduced from the
	symmetries of the compressible Navier-Stokes action.
	In particular, it is shown that the transverse and longitudinal sectors of the velocity field completely decouple, which leads to the usual K41
	results in the limit of low Mach numbers as expected.
	But, more interestingly, the study also reveals that this does not impose that the longitudinal velocity mode behaves trivially.
	Indeed, contrary to previous studies, we show that its behavior is closer to the one of the 3D Burgers equation, which possesses two different
	possible infrared behaviors depending on the relative values of the viscosity and the dispersion of the forcing profile: one which scaling is fixed
	by the Edwards-Wilkinson model, consistent with a $\beta=-2$ scaling, and one corresponding to the Kardar-Parisi-Zhang (KPZ) universality class \cite{Gosteva24}
	remarkably compatible with the solenoidal forcing prediction of \cite{Federrath13}.
	We also highlight, at the very fundamental level, how the symmetries of the action are broken as the Mach number augments, and present the conclusions that
	can be drawn from this, although a complete derivation of the scaling relations in the large Mach number regime will be investigated in further work.

	The article is organised as follows: First, we derive the MSRJD field theory for the compressible Navier-Stokes fluid.
	Then, we derive from the symmetries of the action the different relations between the correlation functions.
	In the next section, the consequences of these relations on the scaling properties of the velocity correlation functions are explained.
	Finally, we conclude.

\section{A field theory for compressible turbulence}

	\subsection{Effective action in the MSRJD formalism}

	In order to proceed, we first need to define an action for our problem, what is done within the MSRJD formalism.
	We mostly adopt the notations of the previous work \cite{Canet15}.
	In a nutshell, this technique allows to build the generating functional of the correlation functions of the velocity field $\mathcal{Z}[\mathbf{J},\overline{\mathbf{J}}
	,K,\overline{K}]$ by the introduction of a velocity response field $\overline{\mathbf{v}}$ and a pressure response field $\overline{p}$.
	Explicitly, this writes:
	\begin{equation}
		\label{eqZ}
		\mathcal{Z}\big[\mathbf{J},\overline{\mathbf{J}},K,\overline{K}\big]=\int\mathcal{D}[\mathbf{v}]\mathcal{D}[\overline{\mathbf{v}}]\mathcal{D}[p]\mathcal{D}
		[\overline{p}]
		e^{-\beta S_{CSNS}[\varphi]-\mathbf{J}\cdot\mathbf{v}-\overline{\mathbf{J}}\cdot\overline{\mathbf{v}}-K\cdot p-\overline{K}\cdot \overline{p}}\,,
	\end{equation}
	where $S_{CSNS}$ is the action of the compressible stochastic Navier-Stokes model, $\varphi$ is a generic notation referring to all the different fields
	in one condensed notation, and it should be understood that the notation for the source terms corresponds to: $\mathbf{J}\cdot\mathbf{v}=\int_x \mathbf{J}(x)
	\cdot\mathbf{v}(x)$, $x$ being the four-vector $(t,\mathbf{x})$, and the shorthand notation is $\int_x=\int d^4x$.
	Using functional derivatives of $\mathcal{Z}$ with respect to the sources, it is possible to generate all the correlation functions of $\mathbf{v}$,
	$\overline{\mathbf{v}}$, $p$ and $\overline{p}$.

	Generally though, $Z$ is not the most useful generating functional.
	Taking its logarithm allows to define the free energy functional $\mathcal{W}$, which is the generator of the connected correlation functions of the velocity and response
	velocity fields:
	\begin{equation}
		\mathbf{W}\big[\mathbf{J},\overline{\mathbf{J}},K,\overline{K}\big] = -k_B T
		\log\left(\mathcal{Z}\big[\mathbf{J},\overline{\mathbf{J}},K,\overline{K}\big]\right)\,.
	\end{equation}
	Going one step further, one can define the Legendre transform of $\mathcal{W}$, $\Gamma$, which is a function of the mean velocity field $\mathbf{u}$
	, the mean response velocity field $\overline{\mathbf{u}}$, the mean pressure field $P$ and the mean response pressure field $\overline{P}$ defined as:
	\begin{equation}
		\begin{split}
			& u_\alpha(x) = \left<v_\alpha(x)\right> = \frac{\delta \mathcal{W}[\varphi]}{\delta J_\alpha(x)}\Bigg|_{U=0} \\
			& \overline{u}_\alpha(x) = \left<\overline{v}_\alpha(x)\right> = \frac{\delta \mathcal{W}[\varphi]}{\delta \overline{J}_\alpha(x)}\Bigg|_{U=0} \\
			& P(x) = \left<p(x)\right> = \frac{\delta \mathcal{W}[\varphi]}{\delta K(x)}\Bigg|_{U=0} \\
			& \overline{P}(x) = \left<\overline{p}(x)\right> = \frac{\delta \mathcal{W}[\varphi]}{\delta \overline{K}(x)}\Bigg|_{U=0}
		\end{split}
	\end{equation}
	where we introduced the notation $U=0$ as a generic term meaning that all the sources are taken equal to 0.

	$\Gamma$ is thus the generator of the cumulant of the velocity, velocity response, pressure and response pressure field distribution functions:
	\begin{equation}
		\Gamma\big[\mathbf{u},\overline{\mathbf{u}}, P, \overline{P}\big] = - \mathcal{W}[U] + \mathbf{J}\cdot\mathbf{u} + \overline{\mathbf{J}}\cdot\overline{\mathbf{u}}
		+ K\cdot P + \overline{K}\cdot\overline{P}\,.
	\end{equation}
	It is also called the \textit{effective action} of the system.

	\subsection{The microscopic action}

		Let us now examine in more details the writing of the action $S_{CSNS}$:
		\begin{equation}
		\label{eqSCSNS}
			\begin{split}
				S_{CSNS}\big[\varphi\big] = \int_{x}& \Bigg\{\overline{\rho}(x)\Big[\partial_t\rho(x)
				+\text{div}\big(\rho(x)\mathbf{v}(x)\big)\Big] \\
									       &+\overline{\mathbf{v}}(x)\cdot\Big[
				\rho(x)\partial_t\mathbf{v}(x) + \rho(x)v_\beta(x)\partial_\beta\mathbf{v}(x)
			-\mathbf{\partial}\cdot\sigma[\mathbf{v}]\Big]\Bigg\}
			\end{split}
		\end{equation}
		Let us first focus on the second line. It corresponds more or less to the usual writing of the stochastic Navier-Stokes (SNS) action, up to three
		minor points: (i) the writing (\ref{eqSCSNS}) involves the stress tensor of the fluid $\sigma$, for reasons that will be clarified later, so that
		the pressure field $p$ is hidden in the writing of $\sigma$ and does not appear explicitly, (ii) there is no response pressure field $\overline{p}$;
		This is a consequence of the fact that in previous studies, only the incompressible version of the SNS action was written, so that a Lagrange multiplier,
		$\overline{p}$, was needed to ensure that the incompressibility property was maintained along the RG flow. However, here we consider a more general,
		compressible flow, so that this Lagrange multiplier is not needed anymore.
		As we are going to show in the following, this does not mean that we will not be able to identify proper quantities for the comparison with the
		incompressible SNS case; (iii) the fluid's density $\rho(x)$ is now a \textit{fluctuating field}, just as the pressure and the velocity.
		Since in the incompressible limit, it becomes equal to a constant, it is not unusual to see it appear in the denominator of the pressure term, rather than
		in the kinetic term of the velocity field. However, this shorthand writing requires an overall redefinition of the action up to a constant, which is
		not possible anymore when $\rho$ fluctuates.

		This leads us to the investigation in more details of the first line of (\ref{eqSCSNS}).
		It is the term enforcing the conservation of the mass in the flowing fluid, which is trivial in the incompressible case, but not anymore when
		$\rho$ fluctuates.
		We thus have to introduce two additional fields with respect to the incompressible case: the density field $\rho(x)$, and its corresponding
		response field $\overline{\rho}(x)$ playing the role of a Lagrange multiplier to ensure matter conservation.
		As a consequence, the generating functional $\mathcal{Z}$ is not a function of $\overline{K}$ anymore, but has two more arguments corresponding to the
		source terms associated with $\rho$ and $\overline{\rho}$.
		In addition, the functional integral now runs over $\rho$ and $\overline{\rho}$ in place of $\overline{p}$ that is not needed anymore.

	\subsection{The stress tensor}

		We are now going to discuss the stress tensor term $\sigma$; But before we proceed, we must introduce some notations that will allow to split
		the different terms appearing in the action as a function of symmetries, which will simplify the expressions in the following.

		As we anticipated earlier, although in our case both compressible and incompressible modes are present in the velocity field, it is interesting to
		separate them.
		This can be done thanks to the longitudinal and transverse projectors, that are defined in the reciprocal space as:
		\begin{equation}
			P_\parallel^{\alpha\beta}(\mathbf{q})=\frac{q_\alpha q_\beta}{q^2}\;,\ \quad P_\perp^{\alpha\beta}(\mathbf{q}) = \delta_{\alpha\beta}
			- P_\parallel^{\alpha\beta}(\mathbf{q})\,;
		\end{equation}
		We can thus define
		\begin{equation}
			\mathbf{v}_\parallel = P_\parallel\cdot\mathbf{v}\;,\ \mathbf{v}_\perp = P_\perp\cdot\mathbf{v}\;,\ \mathbf{v} = \mathbf{v}_\parallel+\mathbf{v}_\perp\,,
		\end{equation}
		where the last equality is a consequence of the fact that $P_\parallel$ and $P_\perp$ are  a complete set of orthogonal projectors.

		Interestingly,
		\begin{equation}
		\label{eqInc}
			\partial_\alpha v_\perp^\alpha = 0\,,
		\end{equation}
		so that $\mathbf{v}_\perp$ can be identified with the incompressible part of the velocity field; $\mathbf{v}_\parallel$ is thus the longitudinal
		mode.
		Note that (\ref{eqInc}) is a direct consequence of the symmetries of $\mathbf{v}_\perp$, no Lagrange multiplier is needed to ensure that
		the RG flow does not break the incompressibility property of $\mathbf{v}_\perp$.

		Let us now turn to the stress tensor itself.
		A natural way of decomposing it is by using the irreducible representations of the SO$(3)$ group; We will thus separate the spin 0,
		diagonal, part of the tensor, $\sigma_0\,\delta_{\alpha\beta}$, and its spin 2 deviatoric part $\Delta\sigma_{\alpha\beta}$.
		In the case of a Newtonian fluid, the deviatoric part depends only on the viscosity of the fluid $\nu$, and the transverse velocity:
		\begin{equation}
			\Delta\sigma_{\alpha\beta}(x) = \nu\, \partial_\alpha v_\perp^\beta(x)\,, 
		\end{equation}
		while the diagonal part involves the pressure field, the second viscosity coefficient $\zeta$, as well as the longitudinal velocity mode:
		\begin{equation}
			\sigma_0(x) = -p(x) + (\nu + 3\zeta)\partial_\alpha v_\parallel^\alpha(x)\,.
		\end{equation}

		To finish the analysis of the stress tensor terms in (\ref{eqSCSNS}), we must comment on the fact that the stress tensor appears in the action only
		through its derivative.
		Because derivatives commute,
		\begin{equation}
			\partial_\parallel \mathbf{v}_\perp(x) = 0 = \partial_\perp \mathbf{v}_\parallel(x)\,.
		\end{equation}
		Also, by definition of the orthogonal projector,
		\begin{equation}
			\mathbf{v}_\perp\cdot \partial = 0 = \overline{\mathbf{v}}_\perp\cdot\partial
		\end{equation}
		(the same decomposition can be applied to any vector field, in particular the velocity response field).

		Consequently, at the level of the microscopic action, the diagonal, spin 0, part of the stress tensor only couples to the longitudinal velocity mode,
		%except for the pressure term, ???????
		while its deviatoric, spin 2, part couples only to the transverse modes.

	\subsection{Mode decoupling}

		Let us now investigate the consequences of the decomposition of the velocity field at the level of the compressible SNS action.

		First, in the density part,
		\begin{equation}
			\text{div}\big(\rho\,\mathbf{v}\big) = \rho \partial_\alpha v_\alpha + v_\alpha \partial_\alpha \rho
			= \rho \partial_\alpha v_\parallel^\alpha + v_\parallel^\alpha \partial_\alpha \rho\,.
		\end{equation}
		Hence, only the longitudinal part of the velocity vector couples to the matter conservation part of the action.

		Second, let us examine the kinetic term:
		\begin{equation}
			\overline{v}_\alpha\cdot\big[\partial_t v_\alpha + v_\beta \partial_\beta v_\alpha \big]
			= \overline{v}_\perp^\alpha\cdot\partial_t v_\perp^\alpha + \overline{v}_\perp^\alpha v_\perp^\beta\partial_\perp^\beta(v_\perp^\beta)
			+ \overline{v}_\parallel^\alpha\cdot\partial_t v_\parallel^\alpha + \overline{v}_\parallel^\alpha 
			v_\parallel^\beta\partial_\parallel^\beta(v_\parallel^\beta)
		\end{equation}
		This term thus fully decouples.

		Lastly, let us analyse the stress tensor term:
		\begin{equation}
			\overline{v}_\alpha \cdot\partial\cdot\sigma_\alpha = 
			\nu \overline{v}_\perp^\alpha\partial^2 v_\perp^\alpha + \overline{v}_\parallel^\alpha \partial_\alpha p+
			\overline{v}_\parallel^\alpha \big[\nu \partial^2 v_\parallel^\alpha + 3\zeta\partial_\alpha \partial_\beta(v_\parallel^\beta)\big]
		\end{equation}
		As for the case of the density fluctuation mode, the transverse response velocity field does not couple to the pressure term.
		This seems to be in contrast to the usual incompressible SNS models, but actually, in this case, the coupling only exists because the
		response pressure field is not prescribed to be transverse in the same way the velocity field does.
		However, the pressure term decouples from the renormalisation group flow anyway \cite{Canet15}, so that this does not bring much change in practice.

		Then, since in $d\geqslant2$ there is always only one longitudinal direction,
		\begin{equation}
			\nu\,\overline{v}_\parallel^\alpha\cdot \partial^2 v_\parallel^\alpha + 3\overline{v}_\parallel^\alpha\zeta\partial_\alpha
			\partial_\beta(v_\parallel^\beta)
			= (\nu + 3\zeta) \overline{v}_\parallel^\alpha\cdot \partial^2 v_\parallel^\alpha
			= \nu_{\parallel} \overline{v}_\parallel^\alpha\cdot \partial^2 v_\parallel^\alpha\,,
		\end{equation}
		where we defined an effective longitudinal viscosity as $\nu_\parallel = \nu + 3\zeta$ at the bare level.
		In the same way, it is useful to define a transverse viscosity, $\nu_\perp = \nu$, which equals the fluid's viscosity at the bare level, but
		gets renormalised independently from $\nu_\parallel$, as different scales are considered, and the fluctuations of the fields renormalise the
		values of these coefficients.

		All in all, there is a full decoupling of the modes in the microscopic action of the compressible SNS model, which should be preserved by the
		RG flow at all scales, so that we can write:
		\begin{equation}
		\label{eqDec}
			S_{CSNS}[\varphi] = S^\parallel_{CSNS}[\varphi_\parallel] + S^\perp_{CSNS}[\varphi_\perp] \ \Rightarrow\ \Gamma[\phi]
			= \Gamma_\parallel[\phi_\parallel] + \Gamma_\perp[\phi_\perp]\,,
		\end{equation}
		with $\phi_\parallel(x) = \left<\varphi_\parallel(x)\right>$, and $\phi_\perp(x)=\left<\varphi_\perp(x)\right>$.

		The transverse action is the same as in the incompressible SNS model, except that the pressure term is absent.
		The longitudinal part is composed of two terms, the matter conservation part, and a term that can be identified with the
		viscous Burgers model, but in 3D and not in 1D (and with an additional pressure term, but as we are going to show, this plays a very minor role).
		This last point reveals to be crucial because, even though the velocity field only has one longitudinal component, the RG fixed point structure
		of the Burgers equation for a field fluctuating in a 1D or 3D environment is significantly different \cite{Gosteva24}.
		It is the first important new element brought by this study.

\section{Symmetries and Ward identities}

	In the context of a RG study in statistical field theory, two types of transformations of the microscopic action are of particular importance: (i) The
	transformations that leave the action invariant, (ii) the transformations for which the end result is linear in one of the microscopic field, and
	can thus be re-absorbed in a redefinition of one of the sources in (\ref{eqZ}).
	In both cases, the full generating functional is left unchanged, which yields exact relations, called Ward identities, relating the correlation functions
	and their generating functionals.

	The study of the impact of such symmetries on the incompressible SNS action can be found in \cite{Canet15}, as for the 3D viscous Burgers equation,
	it is given in \cite{Gosteva24}.
	We will not reproduce the already derived relationships in this work, referring the interested reader to the two references above, but rather concentrate on
	the difference between the already known scenario and the compressible SNS case.
	In order to simplify our study, we can benefit from the decoupling property (\ref{eqDec}) that allows us to study the transverse and longitudinal parts of the
	action separately.

	\subsection{The transverse action}

		We have shown above that the transverse part of the action has the form of the incompressible SNS action without the pressure term.
		This means for example that the pressure gauged symmetry:
		\begin{equation}
			p(x) \mapsto p(x) + \epsilon(x)
		\end{equation}
		that ensured the preservation of the incompressibility property at all scales in the RG flow in the incompressible SNS case \cite{Canet15}
		is not present anymore here.
		This is due to the definition of the transverse velocity field which is incompressible by definition.

		We are not going to make a list of all the other symmetries that are not needed anymore, like the ones involving the pressure response field
		that does not appear, again, because incompressibility is enforced structurally here.

		The major symmetry we have to study is the time-gauged Galilean symmetry,
		\begin{equation}
			\begin{split}
				& v_\perp^\alpha(\mathbf{x},t) \mapsto v_\perp^\alpha(\mathbf{x},t) - \frac{d\epsilon^\alpha}{dt}(t)
				+\epsilon^\beta(t)\partial_\beta v_\perp^\alpha(\mathbf{x},t) \\
				& \overline{v}_\perp^\alpha(\mathbf{x},t) \mapsto \overline{v}_\perp^\alpha(\mathbf{x},t)
				+\epsilon^\beta(t)\partial_\beta\overline{v}_\perp^\alpha(\mathbf{x},t)
			\end{split}
		\end{equation}
		$\epsilon$ being an arbitrary function of time.

		This symmetry is crucial as in the incompressible case, it yields a Ward identity relating the different vertex functions exactly,
		that puts very strong constraints on the RG flow, and in particular driving it towards K41 in the inertial range.
		However here, since $\rho(x)$ has been promoted to a fluctuating field rather than a constant, the action of this transformation on the
		microscopic action $S_{CSNS}^\perp$ is not linear in the fields anymore.

		More precisely, in the limit of low Mach numbers, $|\left<\delta\rho\right>|/\left<\rho\right> \ll 1$,
		so that $\rho$ is a constant to a good degree of approximation
		and the usual Ward identity holds, leading among other things to the Kárman-Howarth relation \cite{Canet15}, and in turn, the K41 scaling.
		However, as the Mach number increases, this property is lost as the Ward identity is broken leaving the scaling of the velocity correlation
		functions, and thus that of the kinetic energy spectrum unconstrained.
		Whether this occurs abruptly in a two scaling fixed points kind of scenario, or more progressively requires a specific RG model to study
		in details the behavior of the RG flow, a task that we reserve to future work since we want to present here only general properties that
		stay valid whatever the truncation scheme devised for actual computations.
		A more precise statement about what happens in the case of moderate to high Mach numbers can still be sorted out, and is presented a little bit further
		in our study.

		All in all, in the limit of low Mach numbers (a precise definition of how low would, again, necessitate a precise RG model), the transverse
		part of the compressible SNS action behaves exactly as its incompressible counterpart.
		However, as the Mach number increases and the relative fluctuations of density becomes bigger, the symmetry that protected the K41 scaling
		in the inertial range vanishes and the behavior of the velocity correlation functions is expected to change.

	\subsection{The longitudinal action}

		The longitudinal action $S_{CSNS}^\parallel$ is composed of two parts: (i) The matter conservation equation part, and (ii) the 3D Burgers
		action with pressure part.

		Before beginning our study, let us remark that the numerous auxiliary fields that are needed to enforce,
		in particular, the preservation of the longitudinal character
		of the velocity field in 3 dimensions, and which are meticulously described in \cite{Gosteva24} are not present anymore, for the same reason
		as above, because this constraint is structurally there already.

		We will thus first begin with the transformation that is absent from the usual Burgers equation, the pressure gauged shift:
		\begin{equation}
			p(x) \mapsto p(x) + \epsilon(x)
		\end{equation}
		Its effect on the action is linear in the longitudinal velocity response field, which leads to the following Ward identity:
		\begin{equation}
			\delta S_{CSNS}^\parallel[\varphi_\parallel]= \int_x \overline{v}_\parallel^\alpha\partial_\alpha \epsilon(x)
			\ \Rightarrow\ \frac{\delta \Gamma_\parallel}{\delta P(x)} = -\partial_\alpha \overline{u}_\parallel^\alpha(x)
			= \left<\frac{\delta S_{CSNS}^\parallel[\varphi_\parallel]}{\delta p(x)}\right>
		\end{equation}
		Namely, the pressure sector is not renormalised.
		This situation is similar to that of the usual incompressible SNS case \cite{Canet15} and should not be misinterpreted: given that there
		exists a non-trivial coupling to the diagonal, spin 0 part of the stress tensor at the level of the action, it cannot be excluded that
		an overpressure is generated within the renormalisation group process, and such a term would be added to the usual pressure term in the
		Burgers equation if one wants to compute the value of the pressure, under a given set of conditions, at a given point.
		But the crucial thing is that such a term would not in any case depend on the pressure that was present in the equation in the first place.
		This conclusion does not depend on the Mach number either since, when $\rho$ is a fluctuating field, its place is in the kinetic term and not
		in front of the pressure term.
		Such a property seems to us to be particularly interesting as in almost all of the previously cited papers dealing with compressible turbulence,
		an equation of state, often of the form $p(x) = c_s^2\,\rho(x)$, $c_s$ being the speed of sound, was used in the derivation of the
		scaling laws, being it by numerical or analytical means.
		What we claim here is that, at least as long as RG analysis is concerned, such an hypothesis is not required at any point in the derivation.
		This simplifies a great deal the initial problem.

		Second, let us examine the effect of a constant shift of the response density field:

		\begin{equation}
			\overline{\rho}(x) \mapsto \overline{\rho}(x)+ \overline{\epsilon}
		\end{equation}

		The variation of the action is~:
		\begin{equation}
			\begin{split}
				\delta S_{CSNS}^\parallel\big[\varphi_\parallel\big] &= \int_{x}\overline{\epsilon}\Big\{
					\partial_t\rho(x) + \rho(x)\partial_\alpha v_\parallel^\alpha(x) + v_\parallel^\alpha(x)\partial_\alpha\rho(x) \Big\} \\
										     &= \int_{x}\overline{\epsilon}\,\partial_t\rho(x)
			\end{split}
		\end{equation}
		where in the last line, we used the fact that the two last terms of the first line are boundary terms.
		Hence, the non-linear term cancels, and the resulting transformation is again linear in the fields, which allows us to write,
		\begin{equation}
			\frac{\delta \Gamma_\parallel}{\partial \overline{R}} = \left<\int_{x} \partial_t\rho(x)\right> = \left<\frac{\delta S_{CSNS}^\parallel}
			{\partial \overline{\rho}}\right>
		\end{equation}
		(where $\overline{R} = \left<\overline{\rho}\right>$).

		The physical interpretation of this result is quite transparent~: The mass conservation equation is enforced at all scales
		in the renormalisation group flow.
		But, the above equation also means that the mass conservation part of the effective action effectively decouples from
		the renormalisation group flow.
		Putting all these results together, we get that the renormalisation of the longitudinal part of the effective action
		is the same as that of the 3D Burgers model (since all the new terms decouple from the flow).
		This is a major simplification compared to the initial problem.

		It should be understood though in the above comment that $\rho(x)$ is nonetheless still a fluctuating field.
		The fact that the mass conservation part is not coupled to the general flow means that the renormalisation of $\rho$
		is entirely determined as a function of that of $v_\parallel^\alpha$, not that the density field is not renormalised.
		The fact that the density field still fluctuates is of crucial importance, as we can already anticipate from the study of the transverse
		part of the action.
		It is also required for our results to be in agreement with the previous literature presented in the introduction: Indeed, if the longitudinal mode
		was the only acoustic mode, the study of Kraichnan \cite{Kraichnan54} would imply that the longitudinal velocity field alone carries the
		same amount of energy as the two transverse modes, which would also forbid scaling laws for the power spectrum that would be in agreement with
		the values found in previous studies.

		Finally, a crucial symmetry is of course the time-gauged Galilean symmetry, but we are not going to reproduce the results here as they are exactly the
		same as in the incompressible case: In the limit of low Mach numbers, the symmetry behaves exactly as in the 3D Burgers model (see \cite{Gosteva24}),
		and the scaling properties follow; As the Mach number increases however, the relative fluctuation of $\rho$ compared to its mean value is not
		negligible anymore, and the action of the time-gauged Galilean symmetry is not linear in the fluctuating fields anymore, which means that its action
		cannot be cancelled by a redefinition of the source terms.
		The importance of this symmetry should absolutely not be understated though, as it provides the scaling in the low Mach number limit.
		Indeed, as we are going to show, it enforces a scaling for the longitudinal mode that is different from that of the transverse modes,
		a result which supports a type of scenario like the one presented in \cite{Federrath13} with two types of modes with two types of scaling
		both contributing to the power spectrum of the kinetic energy dissipation.

	\subsection{Going further with bilinear sources}

		Even restraining ourselves to exact statements true whatever the applied RG procedure, it is a bit frustrating not to have a Ward identity
		corresponding to the time-gauged Galilean symmetry in the case of moderate to high Mach numbers.
		As it turns out, this Ward identity provides a hierarchy of relationships between the successive vertex operators (the functional derivatives
		of $\Gamma$ with respect of the different fields involved) which is the backbone of the further RG analysis.

		There is one way to go further than what we have done so far, which consists in adding to the definition of $\mathcal{Z}$ two new source
		terms, $\mathbf{L}$ and $\overline{\mathbf{L}}$, that instead of coupling \textit{linearly} to the fields couple to a composite operator consisting in two
		fields evaluated at the same point.
		Explicitly, we add $\rho\cdot \mathbf{L}\cdot\mathbf{v}$ and $\rho\cdot\overline{\mathbf{L}}\cdot\overline{\mathbf{v}}$.
		These two terms are not to be taken into account in the Legendre transform that defines $\Gamma$ from $\mathcal{W}$, so that the previous identities
		are not modified.
		However, it allows to provide a form where the action of the time-gauged Galilean symmetry, augmented with the transformation of the density field
		\begin{equation}
			\rho(x)\mapsto \rho(x) + \epsilon_\alpha(t)\partial_\alpha \rho(x)\,,
		\end{equation}
		can be re-absorbed in a redefinition of the bilinear sources, finally leading to the following Ward identity:
		\begin{equation}
		\label{eqW}
			\begin{split}
				\int_{x,y,z}&\Bigg\{\Big(\delta_{\alpha\beta}\partial_t + \partial_\beta u_\alpha(x)\Big)\frac{\delta^2\Gamma}{\delta u(y)\delta\rho(z)}
				+\partial_\beta\overline{u}_\alpha(x)\frac{\delta^2\Gamma}{\delta \overline{u}_\alpha(y)\delta\rho(z)} \Bigg\} \\
					    &= -\left<\int_x \rho(x)\partial_t^2\overline{v}_\beta(x) + \partial_\alpha\big(\rho(x)\big)\overline{v}_\beta(x)
					    \partial_t v_\beta(x) + \partial_\alpha\big(\rho(x)\big)\overline{v}_\gamma(x) v_\beta(x) \partial_\beta v_\gamma(x)\right>
			\end{split}
		\end{equation}
		It is important to note that the introduction of the bilinear source operators did not technically break the decoupling property, that can still be shown
		to exist in the new formalism, but it restored a coupling between the density mode and the transverse velocity fluctuations that was not present in the
		low Mach number limit.
		We insist again on the fact that it is not a fundamental breaking of the RG flow properties since we have shown that, even in the supersonic regime,
		the flow of the density field completely decouples, but the fact that the density field, which fluctuates in a way fixed by the renormalisation
		of the longitudinal component of the velocity, enters the fundamental equation at the basis of all the hierarchies between vertex operators
		even for the transverse modes shows how profound the consequences of going from the subsonic to the supersonic regime go.

		It can be checked that (\ref{eqW}) yields back the usual Ward identities derived in \cite{Canet15} and \cite{Gosteva24} in the limit
		where $\rho$ becomes a constant.

\section{Consequences on scaling}

	First of all, let us mention that the decoupling property holds irrespective of the Mach number and the values of the scale in the RG flow, so that
	we can write at all scales than:
	\begin{equation}
		\mathcal{C}_{\perp\parallel}(x) = \left<\mathbf{v}_\perp(x)\cdot\mathbf{v}_\parallel(\mathbf{0},0)\right> =
		\mathcal{C}_{\parallel\perp}(x) = \left<\mathbf{v}_\parallel(x)\cdot\mathbf{v}_\perp(\mathbf{0},0)\right> = 0
	\end{equation}
	All the non trivial contributions to the energy power spectrum thus come from coupling between two transverse or two longitudinal modes.

	\subsection{Moderately subsonic flows}

		In the case of moderately subsonic flows, we have shown that the weakness of the relative fluctuations of the density mode relative to its mean
		implies that the transverse velocity modes behave with the K41 scaling.
		It is thus predicted that in this region, the contribution of the transverse modes to the power spectrum decays in the inertial range with
		$\beta_{K41}=-5/3$.

\begin{figure}[h]
			\begin{center}
				\includegraphics[scale=0.7]{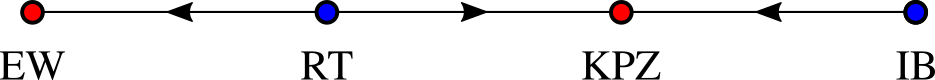}
			\end{center}
		\caption{Schematic picture of the basins of attraction of the various fixed points involved in the renormalisation group flow of the
		viscous Burgers equation: Edwards-Wilkinson (EW), roughening transition (RT), Kardar-Parisi-Zhang (KPZ) and inviscid Burgers (IB).
		The red points correspond to the infrared scaling, the blue ones to the ultraviolet scaling.
		The arrows indicate the direction of the RG flow.}
		\label{figF}
		\end{figure}

		For the longitudinal modes however, the scenario is a bit more subtle, as it has been shown in \cite{Gosteva24}.
		Indeed, depending on the relative values of the bare viscosity of the fluid and the dispersion of the random forcing, different scenarios are
		possible (see Fig.~\ref{figF}):
		(i) For the lowest viscosity to force dispersion ratio, the large scale, infrared (IR) behavior is given by the Edwards-Wilkinson fixed point,
		while the small scale ultraviolet (UV) behavior is given by the class of universality of the roughening transition; (ii) beyond a first critical value
		the IR behaviour is given by the KPZ fixed point instead; (iii) For the highest values of the bare ratio, the IR behavior is still in the KPZ
		basin of attraction, but the UV behavior is given by the inviscid Burgers fixed point.

		Let us underline here the crucial role played by the fact that, although it has only one direction, the longitudinal mode behaves according to the
		3D, and not the usual 1D, Burgers equation.
		This property is key in guaranteeing the richness of the RG flow diagram of Fig.~\ref{figF}.

		Putting everything together, using the values given in \cite{Gosteva24} obtained in a simplified model, we can get the following rough values for the
		critical exponent $\beta$ corresponding to the contribution of the longitudinal mode: $\beta_{EW}=-2$, $\beta_{RT} = -1$, $\beta_{KPZ} = -1.5$,
		$\beta_{IB}=-2$.
		These values must of course be treated more carefully before any quantitative comparison with numerical results are made, but in absence of a full
		RG treatment of our model, let us sketch a qualitative scenario.

		Near the Kolmogorov scale where the dissipation occurs, the UV scaling dominates. In this regime, $\beta_{IB}<\beta_{K41}<\beta_{RT}$.
		Thus, depending on the viscosity of the fluid, the UV part of the kinetic energy power spectrum is either dominated by the incompressible K41
		scaling, or the RT scaling.
		In the limit of incompressible fluids, the forcing is taken to be purely solenoidal, in which case the longitudinal modes are not excited,
		and the scaling is given by its K41 value, in agreement with expectations.

		At the other end of the inertial range, it is expected that the IR behavior dominates.
		Here, we have $\beta_{EW}<\beta_{K41}<\beta_{KPZ}$.
		So, depending again on the conditions at the bare level, it is expected that either the EW or the K41 fixed point dominates the IR regime.

		Now, this qualitative sketch needs to be nuanced quite a lot.
		First, the presented picture supposes that the inertial range is such that its low wave number end lies in the IR basin of attraction of the RG
		flow, and its high wave number end is in the UV basin of attraction.
		This is not a given.
		Then, as stated above, the values of $\beta$ given here are quite rough.
		In particular, that of the KPZ fixed point, that lies dangerously close to the value of the K41 fixed point, would require a more careful estimate
		since if its value is a little bit lower, the longitudinal mode could dominate the energy power spectrum in the IR limit whatever the initial conditions
		of the RG flow, the variation in the value of $\beta$ being attributed to the change from the EW to the KPZ basin of attraction.

		Finally, let us comment briefly on the results of the study \cite{Federrath13}, setting apart the high value of the chosen Mach number (as explained
		above, a careful analysis of how the transition from low relative density fluctuations to high relative density fluctuations occurs in our model
		would require further work).
		If the forcing is purely compressible, only the longitudinal mode is excited, so that the observed behavior in the limit of low viscosity is given
		by the EW fixed point with $\beta=-2$.
		If the forcing is purely solenoidal, it is expected that only the transverse mode is excited, so that $\beta\simeq - 1.67$.
		A qualitative scenario of this type would thus be in qualitative agreement with our model.
		Obviously, one should not overstate the value of the result above, a number of nuances having been already expressed above.

	\subsection{The limit of moderate to high Mach numbers}

		In the limit of moderate to high Mach numbers, or more precisely as far as our model is concerned, the limit of high relative density fluctuations,
		the profound modification of the Ward identity (\ref{eqW}) will certainly modify the qualitative picture exposed for the low Mach numbers.
		However, in absence of already established results in such a model, we cannot provide scaling predictions for the energy power spectrum.
		Moreover, if in the case of low Mach numbers, the evolution of the kinetic energy power spectrum can be, at least naively, extrapolated
		from results concerning the velocity correlation function, it is not so anymore when the fluctuations of $\rho$ become important,
		and the kinetic energy reveals itself to be a three-point and not two-point correlation function anymore.
		Said otherwise, the renormalisation of the density field, which enters the definition of the kinetic energy, has to be taken into account on
		top of the renormalisation of the velocity field.
		This point further obscures the relation between the low and high Mach number behaviors.

		Without entering too much into the details of such a computation, a first scaling analysis can still be performed, to yield relations between the
		different critical exponents.
		Remembering that the density conservation equation only appears in the longitudinal part of the action, we can define the following
		anomalous dimensions for a generic quantity $X$:
		\begin{equation}
			\eta_X=k\partial_k\big(\log(X_k)\big)
		\end{equation}
		where $X_k$ is the value of $X$ corrected for all the fluctuations at scales $q\geqslant k$; In that way we define $\eta_\nu$
		from $\nu_\parallel$ (remember that the transverse viscosity renormalises independently), $\eta_\rho$ and $\eta_{\overline{\rho}}$.
		Then, we define the dimension $d\geqslant2$, the dynamical exponent $z$, and the dimensions of the (longitudinal)
		velocity fields $\Delta(v)$ and $\Delta(\overline{v})$.
		Standard RG derivation leads to the following relations:
		\begin{equation}
			\begin{split}
				& \Delta(\overline{v}) = d - 1 + \eta_\nu \\
				& \Delta(v) = 1 - \eta_\nu + \eta_\rho \\
				& z = 2 - \eta_\nu + \eta_\rho \\
				& \eta_{\overline{\rho}} + \eta_\rho + d = 0
			\end{split}
		\end{equation}

		It can be checked that these equations are compatible with the low Mach number model in the limit where $\rho$ behaves as a constant:
		$\eta_\rho\rightarrow0$, $\eta_{\overline{\rho}}\rightarrow -d$.

\section{Conclusion}

	All in all, we have presented in this article a comprehensive framework that allows to derive exact relations between the correlation functions
	of fluctuating fields in compressible turbulence.
	Although a detailed analysis, within a well-defined RG model (that would introduce approximations in the model) is necessary to be able to clarify
	once and for all the unsolved problems such as the exponents of the compressible turbulence energy power spectrum, at least in the inertial range,
	or the type of transition that occurs between the low Mach number and high Mach number regimes, a number of lessons can still be drawn from
	this exact derivation.
	Instead of stating again what has been derived in the above, let us conclude by stressing what is, in our opinion, the most important of these lessons:
	The longitudinal velocity mode evolves according to a 3D, and not the usual 1D, Burgers equation model, which opens the way to a much richer RG flow
	diagram, corresponding to a diversity of possible values for the critical exponent $\beta$, already in the low Mach number regime.
	This surely plays a role in the diversity of results observed in numerical simulations, and should be taken into account in the analysis of the
	forthcoming studies.

\section*{Acknowledgements}

	I am deeply thankful to L. Canet and L. Gosteva whose previous work has provided me with the inspiration for the present study.
	I am also grateful to C. Federrath for kindly accepting to share his knowledge about simulations in compressible turbulent flows.

\bibliography{CT.bib}

\end{document}